\documentclass{PoS}

\title{Particles and the Universe}

\ShortTitle{Particles and the Universe}

\author{\speaker{George Lazarides}
\\
School of Electrical and Computer Engineering, Faculty of 
Engineering, Aristotle University of Thessaloniki, Thessaloniki 
54124, Greece\\
E-mail: \email{lazaride@eng.auth.gr}}

\abstract{The early stages of the universe evolution are discussed
according to the hot big bang model and the grand unified
theories. The shortcomings of big bang are summarized and their
resolution by inflationary cosmology is sketched. Cosmological
inflation, the subsequent oscillation and decay of the inflaton
field, and the resulting reheating of the universe are studied
in some detail. The density perturbations produced by inflation
and the temperature fluctuations of the cosmic microwave
background radiation are introduced. Baryogenesis via non-thermal
leptogenesis is analyzed and dark energy and matter in the 
universe are presented. Quantum gravity and string theory are 
very briefly introduced. The problem of initial conditions for
inflation is discussed in the light of string theory and the 
possibly detectable primordial gravity waves from inflation are 
mentioned.}

\FullConference{Corfu Summer Institute 2019 "School and 
Workshops on Elementary Particle Physics and Gravity" 
(CORFU2019)\\
31 August - 25 September 2019\\
Corfu, Greece}

\begin{document}

\section{Introduction}

The observed Hubble expansion of the universe, the 
discovery of the cosmic microwave background radiation 
(CMBR) and the success of nucleosynthesis (see e.g. 
Ref.~\cite{bbn}) in reproducing the observed abundance 
of light elements in the universe had established the 
standard big bang (SBB) cosmological model (for a 
textbook treatment, see e.g. Ref.~\cite{wkt}). This 
model together with the grand unified theories (GUTs) 
\cite{ggps} of strong, weak, and electromagnetic 
interactions provides the scientific framework for 
studying the early stages of the universe.

Despite its great successes, the SBB model had some 
shortcomings. One of them is the so-called horizon 
problem. The CMBR received now has been emitted from 
regions of the universe which, according to the SBB 
model, had never communicated before sending this
radiation to us. The question then arises how come 
the temperature of this radiation from these regions 
is so finely tuned as the measurements of the COBE 
\cite{cobe}, the WMAP \cite{wmap}, and the 
Planck \cite{planck} satellites show. Another 
important puzzle is the so-called flatness problem. 
It is a fact \cite{planck} that the present universe 
appears to be very flat. This means that, in its 
early stages, the universe must have been flat with 
a great accuracy, which requires an extreme fine 
tuning of its initial conditions. Also, combined 
with GUTs which predict the existence of superheavy 
magnetic monopoles \cite{monopole}, the SBB model 
leads \cite{preskill} to a catastrophic overproduction 
of these monopoles. Finally, the model has no 
explanation for the small density perturbations 
required for explaining the structure formation in the 
universe (for a pedagogical discussion, see e.g. 
Ref.~\cite{structure}) and the generation of the 
observed \cite{cobe,wmap,planck} temperature 
fluctuations in the CMBR.

Cosmological inflation \cite{guth} offers a solution to 
all these problems of the SBB model (for a textbook 
introductions or reviews on inflation, see e.g.
Ref.~\cite{bookinflation}). The idea is that, in the 
early universe, a real scalar field (the inflaton) was 
displaced from its vacuum value. If the potential of 
this field is quite flat, the roll-over of the field 
towards the vacuum can be very slow for a period of 
time during which the energy density of the universe 
is dominated by the 
almost constant potential energy density of the inflaton. 
Consequently, the universe undergoes a period of 
quasi-exponential expansion, which can readily solve the 
horizon and flatness problems by stretching the distance 
over which causal contact is established and reducing
any pre-existing curvature in the universe. It can also 
adequately dilute the GUT magnetic monopoles and provide 
us with the primordial density perturbations which are 
needed for explaining the large scale structure in the 
universe \cite{structure} and the temperature fluctuations 
observed in the CMBR. Inflation can occur during the GUT
phase transition at which the GUT gauge symmetry breaks by 
the vacuum expectation value (VEV) of a Higgs field, which 
also plays the role of the inflaton.

After the end of inflation, the inflaton starts performing 
damped oscillations about the vacuum and decays into light
particles. The resulting radiation energy density eventually
dominates over the field energy density and the universe
returns to a normal big bang type evolution. The temperature
at which this occurs is historically called reheat 
temperature although there is neither supercooling nor
reheating of the universe \cite{reheat} (see also
Ref.~\cite{dilution}).

An acceptable inflationary scenario must necessarily be 
followed by a successful reheating process which, in the 
case of supersymmetry, must satisfy the gravitino 
constraint \cite{gravitino} on the reheat temperature,
$T_{\rm r}\lesssim 10^9~{\rm GeV}$. This process must also
generate the observed baryon asymmetry of the universe. 
In inflationary models, it is generally preferable to 
generate the baryon asymmetry by non-thermal 
\cite{nonthermalepto} leptogenesis \cite{thermallepto}, 
i.e. by first producing a primordial lepton asymmetry 
from the decay products of the inflaton. This asymmetry 
is then partly converted into baryon asymmetry by the 
non-perturbative electroweak sphaleron effects 
\cite{dimopoulos,sphaleron}. Actually, in many specific
models, this is the only way to generate the baryon 
asymmetry of the universe since the inflaton  decays into 
right handed neutrinos. The subsequent decay of these 
fields into a lepton (antilepton) and an electroweak
Higgs field can only produce a lepton asymmetry.

Recent measurements \cite{planck} confirmed the prediction 
of inflation that the present universe is flat. They also 
reveal that matter constitutes only $27\%$ of the universe. 
The rest $73\%$ of the universe is in the form of dark 
energy, i.e. in a form close to a cosmological constant. 
This means that this energy is basically not diluted by the 
expansion of the universe exactly as the energy driving 
inflation. On the other hand, the baryonic (visible) matter 
constitutes only $4.85\%$ of the universe as 
nucleosynthesis \cite{bbn} and the Planck satellite 
\cite{planck} have shown. Consequently, about $22\%$ 
of the universe is made of some form of dark matter. The 
most promising candidate for dark matter \cite{LSP} is the 
lightest supersymmetric particle (LSP) since it is protected 
by a discrete $Z_2$ R-parity symmetry from decaying into 
lighter particles. The LSPs can annihilate in pairs or 
coannihilate with the next-to-LSPs. Their relic density can 
be reduced to the observed value of dark matter abundance 
mainly by resonant pair annihilation \cite{Apole} or 
coannihilation \cite{coannih} with next-to-LSPs which 
happened to be almost degenerate with them. Other possible 
dark matter candidates include the axion \cite{PecceiQuinn} 
and superheavy \cite{wimpzilla} or intermediate scale mass
\cite{interDM,inter2} fermions.

At cosmic times greater than the Planck time 
$t_{\rm P}\sim 10^{-43}~{\rm sec}$, gravity is 
adequately described  by the classical theory of general 
relativity. However, for smaller cosmic times, quantum
fluctuations of gravity are present. Therefore, it is 
desirable to have a quantum theory of gravity and if
possible unify gravity with the other three interactions, 
which are described by relativistic quantum field theory, 
so as to obtain the theory of everything. However, in 
contrast with the other three interactions, the quantum 
field 
theory for gravity is not renormalizable. This means that
the infinities appearing in the various calculations and
originating from the point-like character of particles 
cannot be systematically gathered in a finite number of 
parameters. Consequently, the theory is lacking 
predictability. The theory of (super)strings 
\cite{superstrings} was proposed in order to cure this 
difficulty. The idea is that the fundamental objects are 
not point-like particles but one dimensional strings of 
size $\ell_{\rm P}\sim 10^{-33}~{\rm cm}$. Their various 
vibrational modes, one of which is the graviton, appear as 
particles with different quantum numbers. The main 
disadvantage of string theory is that it admits a huge 
number of solutions ($\sim 10^{500}$), but none of them 
reproduces exactly our universe. It predicts that there 
exist ten spacetime dimensions (or eleven in the case of 
M theory). Six of them are compactified to form a 
6-dimensional compact space of size $\ell_{\rm P}$, 
while the other four dimensions 
remain open and are the usual spacetime dimensions. The 
geometric structure of the compactified dimensions 
determines many of the phenomena in the 4-dimensional 
spacetime.

One problem we can address in the light of string theory 
is the problem of initial conditions \cite{initial} for 
inflation which takes place at the GUT transition at 
$t\sim 10^{-37}~{\rm sec} \gg t_{\rm P}$. At the onset 
of inflation, one needs a large homogeneous 
region around the GUT phase transition. However, this 
region consists of many smaller homogeneous regions 
originating from the Planck era, where causal communication 
is restricted to distances of the order of the Planck 
length. The resolution of this problem 
requires a first stage of inflation near or before the 
big bang, which provides the necessary homogenization for 
the onset of conventional inflation. Such a primordial 
inflation can take place in the pre-big-bang period. During 
the motion of the universe towards the initial singularity, 
we have conditions of very high curvature and the extra 
dimensions contract and compactify. This causes inflation 
in the four open dimensions. Another application of quantum 
gravity could be the explanation of the origin of primordial 
gravity waves if such waves are detected in the future. The 
Planck satellite measurements \cite{planck} imply an upper 
bound on the tensor-to-scalar ratio $r\lesssim 0.06$. 
Therefore, one cannot exclude that future experiments may 
detect primordial gravity waves originating from inflation. 
Quantum gravity may then be required to understand the origin 
of these waves.

In Sec.~\ref{sec:bbc}, we summarize the salient features of 
SBB cosmology, while in Sec.~\ref{sec:trans} 
we sketch the series of phase transitions the universe 
underwent in accordance with GUTs. The shortcomings of big 
bang and their resolution by inflation are discussed in 
Secs.~\ref{sec:puzzles} and \ref{sec:inflation}, respectively. 
Sec.~\ref{sec:detail} is devoted to the detailed discussion
of inflation, reheating, and density and temperature 
fluctuations. Baryogenesis via leptogenesis is the subject
of Sec.~\ref{sec:baryo}. In Secs.~\ref{sec:DE} and 
\ref{sec:DM}, we review the composition of the universe, 
while in Secs.~\ref{sec:qg} and \ref{sec:string} we briefly
refer to quantum gravity and string theory with some possible
applications in cosmology. Finally, in Sec.~\ref{sec:concl}
we summarize our conclusions.      

\section{Big Bang Cosmology}
\label{sec:bbc}

We start with an introduction to the SBB model \cite{wkt}. 
For cosmic times $t\gtrsim t_{\rm P}\equiv m_{\rm P}^{-1}
\sim 10^{-43}~{\rm{sec}}$ 
($m_{\rm P}\simeq 2.44\times 10^{18}~{\rm{GeV}}$ is 
the reduced Planck scale) after the big bang, the quantum 
fluctuations of gravity cease to exist and gravitation can 
be described by the classical theory of general relativity. 
Strong, weak, and electromagnetic interactions, however, 
are described by gauge theories which are relativistic 
quantum field theories.

An important starting point is the so-called {\it cosmological 
principle}, which states that, at large scales, the universe 
is homogeneous and isotropic. The strongest evidence for this  
is the observed \cite{cobe,wmap,planck} isotropy of the CMBR. 
The spacetime metric then takes the Robertson-Walker form
\begin{equation}
ds^{2}=-dt^{2}+ a^{2}(t)\left[\frac{dr^{2}}
{1-kr^2}+r^{2}(d\vartheta^{2}+\sin^{2}\vartheta~ d\varphi^{2})
\right],
\label{eq:rw}
\end{equation}
where $r$, $\varphi$, and $\vartheta$ are {\it comoving} polar
coordinates, remaining fixed for objects that just follow 
the cosmological expansion. The parameter $k$ is the 
scalar curvature of the 3-space and $k=0$, $>0$, or $<0$ 
corresponds to flat, closed, or open universe. The 
dimensionless parameter $a(t)$ is the scale factor of the 
universe. It is normalized so that $a_{0}\equiv a(t_{0})=1$, 
where $t_{0}$ is the present cosmic time.

The instantaneous radial physical distance is given by
\begin{equation}
R=a(t)\int_{0}^{r}\frac{dr}{(1-kr^{2})^\frac{1}{2}}.
\label{eq:dist}
\end{equation}
For flat universe ($k=0$), $\bar{R}=a(t)\bar{r}$ 
($\bar{r}$ is a comoving and  $\bar{R}$ a physical radial 
vector in 3-space) and the velocity of a comoving object is 
\begin{equation}
\bar{V}=\frac{\dot{a}}{a}\bar{R}\equiv H(t)\bar{R},
\label{eq:hubblelaw}
\end{equation}
where the overdot denotes derivation with respect to $t$ 
and $H(t)$ is the Hubble parameter. This equation is the 
well-known Hubble law asserting that all objects run away 
from each other with velocities proportional to their 
distances. This law is the first success of SBB cosmology.

Energy-momentum conservation yields the continuity equation
\begin{equation}
\frac{d\rho}{dt}=-3H(t)(\rho+p),
\label{eq:continuity}
\end{equation}
where $\rho$ and $p$ are the energy density and pressure in 
the universe. The first term in the right-hand side (RHS) 
of this equation describes the dilution of the energy due 
to the Hubble expansion and the second term the work done 
by pressure. Einstein's equations for the Robertson-Walker 
metric take the form of the Friedmann equation
\begin{equation}
H^{2}\equiv \left(\frac{\dot{a}(t)}{a(t)}\right)^{2} =
\frac{8\pi G}{3}\rho-\frac{k}{a^{2}},
\label{eq:friedmann}
\end{equation}
where $G$ is Newton's gravitational constant. Averaging 
the pressure $p$, we write $\rho+p=\gamma\rho$ and 
Eq.~(\ref{eq:continuity}) gives $\rho\propto
a^{-3\gamma}$. From Eq.~(\ref{eq:friedmann}) with $k=0$, 
we then get $a(t)\propto t^{2/3\gamma}$. For a universe 
dominated by pressureless matter, $\gamma=1$ and, thus, 
$\rho\propto a^{-3}$ and $a(t)=(t/t_{0})^{2/3}$. This is 
interpreted as mere dilution of a fixed number of particles 
in a comoving volume due to the Hubble expansion. For a 
radiation dominated universe, $p=\rho/3$ and, 
thus, $\gamma=4/3$, which gives $\rho\propto a^{-4}$ and 
$a(t)=(t/t_{0})^{1/2}$. The extra factor of $a(t)$ in the 
energy dilution is due to the red-shifting of all wave 
lengths by the Hubble expansion. 

The early universe is radiation dominated and its energy 
density is given by 
\begin{equation}
\rho=\frac{\pi^{2}}{30}g_*~T^{4},
\label{eq:boltzman}
\end{equation}
where $T$ is the cosmic temperature and $g_{*}=N_{b}+(7/8)
N_{f}$ is the effective number of massless degrees of 
freedom with $N_{b(f)}$ being the number of massless 
bosonic (fermionic) degrees of freedom. The 
temperature-time relation during radiation dominance is 
derived from Eq.~(\ref{eq:friedmann})
(with $k=0$):
\begin{equation}
T^{2}=\frac{3\sqrt{5}m_{\rm P}}{\sqrt{2g_*}\,\pi\, t},
\quad m_{\rm P}\equiv (8\pi G)^{-\frac{1}{2}}= 
{\rm the~reduced~Planck~mass}.
\label{eq:temptime}
\end{equation}
Classically, the expansion starts at $t=0$ with 
$T=\infty$ and $a=0$. This initial singularity is, 
however, not physical since general relativity fails 
for $t\lesssim t_{\rm P}$. The only 
meaningful statement is that the universe, after a 
yet unknown initial stage, emerges at $t\sim t_{\rm P}$ 
with $T\sim m_{\rm P}$.

An important notion is the notion of particle horizon 
$d_{H}(t)$, which is the instantaneous distance at $t$ 
traveled by light since the beginning of time ($t=0$). 
From Eq.~(\ref{eq:rw}), we find that 
\begin{equation}
d_{H}(t)=a(t)\int_{0}^{t} \frac{dt^{\prime}}
{a(t^{\prime})},
\label{eq:hor}
\end{equation}
which is finite and coincides with the size of the 
universe already seen at time $t$ or, equivalently, with 
the distance over which causal contact has been 
established at $t$. For matter (radiation) dominated 
universe, we have $d_{H}(t)=2H^{-1}(t)=3t$ ($d_{H}(t)=
H^{-1}(t)=2t$). After the Planck satellite measurements 
\cite{planck}, the present age of our universe is
estimated to be $t_{0}\simeq (13.801\pm 0.024)\times
10^9~{\rm years}$, the present value of the Hubble 
parameter $H_0=100\,h~{\rm km\,sec^{-1}\,Mpc^{-1}}$ with 
$h\simeq 0.674\pm 0.005$, and the present critical density 
corresponding to a flat universe $\rho_{\rm c}=3H_{0}^{2}/ 
8\pi G\simeq 0.86\times 10^{-29}~{\rm{gm/cm^{3}}}$. The 
fraction of the actual to the critical density is $\Omega
\equiv\rho/\rho_{\rm c}\simeq 1\pm 0.01$, which means that 
our present universe is very flat. 

\section{Phase Transitions in the Universe}
\label{sec:trans}

GUTs together with the SBB model predict that, as the 
universe expands and cools after the big bang, it 
undergoes \cite{kl} a series of phase transitions 
during which the GUT gauge symmetry is gradually 
reduced and several important phenomena take place. 
For definiteness, we consider here a simple 
non-supersymmetric $SU(5)$ GUT model, but the 
discussion can be readily extended to include 
other gauge groups such as $E_6$, $SO(10)$, 
$SU(4)_{\rm c}\times SU(2)_{\rm L}\times 
SU(2)_{\rm R}$, and $SU(3)^{3}$ \cite{trinification} 
with or without supersymmetry. At a scale $M_{\rm G}
\sim 10^{16}~{\rm{GeV}}$ (the GUT mass scale), 
$SU(5)$ breaks to the standard model gauge group 
$G_{\rm SM}=SU(3)_{\rm c}\times SU(2)_{\rm L}
\times U(1)_Y$ by the VEV of an appropriate Higgs 
field $\phi$. Subsequently, $G_{\rm SM}$ is broken to 
$SU(3)_{\rm c}\times U(1)_{\rm em}$ at the electroweak 
scale $M_{\rm W}$ (${\rm SU}(3)_{\rm c}$ and 
${\rm U}(1)_{\rm em}$ are, respectively, the gauge 
groups of strong and electromagnetic interactions).

Initially, $SU(5)$ was unbroken and the universe was 
filled with a hot soup of massless particles including 
photons, quarks, leptons, gluons, weak gauge bosons 
$W^{\pm}$, $Z^{0}$, 
GUT gauge bosons $X$, $Y$, and several Higgs bosons. 
At $t\sim 10^{-37}~{\rm{sec}}$ ($T\sim 10^{16}~
{\rm{GeV}}$), $SU(5)$ broke down to $G_{\rm SM}$ and 
the $X$, $Y$ bosons and some Higgs bosons acquired masses 
$\sim M_{\rm G}$. Their out-of-equilibrium decay could, in 
principle, generate \cite{dimopoulos,bau} the observed 
baryon asymmetry of the universe, i.e. an excess of 
baryons over antibaryons. Important ingredients
are the violation of baryon number, which is inherent 
in GUTs, and C and CP violation. This is the second 
(potential) success of the SBB model.

Moreover, at the GUT phase transition, topologically 
stable extended objects \cite{kibble} such as magnetic 
monopoles \cite{monopole}, cosmic strings \cite{string}, 
or domain walls \cite{wall} can also be generated. 
Monopoles, which exist in most GUTs, can lead into
problems \cite{preskill} which are, however, avoided by 
inflation \cite{guth,bookinflation} (see below). This 
is a period of exponentially fast expansion of the 
universe which can occur during the GUT phase transition 
and can totally remove the monopoles from the scene.
Alternatively, a more moderate inflation such as thermal 
inflation \cite{thermalinf}, associated with a phase 
transition occurring at a temperature of the order of the 
electroweak scale, can dilute them to an acceptable, but 
possibly measurable level. Cosmic strings from GUTs 
\cite{cosmicstring}, on the other hand, can generate 
gravity waves \cite{gws}, which will be possibly measurable 
\cite{interDM,gwsstrings,stringsmono} by future experiments. 
Finally, domain 
walls are \cite{wall} catastrophic and GUTs should
be constructed so that they avoid them (see e.g. 
Ref.~\cite{axion}) or inflation should extinguish them. 
Note that, in some cases, more complex extended objects 
such as walls bounded by strings \cite{wallsbounded} or 
strings connecting monopoles \cite{stringsbounded} can be 
temporarily produced.

At $t\sim 10^{-10}~{\rm{sec}}$ ($T\sim 100 ~{\rm{GeV}}$), 
the electroweak phase transition takes place and $G_{\rm SM}$ 
breaks to ${\rm SU}(3)_{\rm c}\times {\rm U}(1)_{\rm em}$. 
The electroweak Higgs field as well as the weak gauge bosons 
$W^{\pm}$, $Z^{0}$ acquire 
masses $\sim M_{\rm W}$. Subsequently, at $t\sim 10^{-4}~
{\rm{sec}}$ ($T\sim 1~{\rm{GeV}}$), color is confined and 
the quarks come together to form hadrons. The direct 
involvement of particle 
physics ends here. We will, however, sketch some of the
subsequent cosmological events since they provide crucial 
information on the early universe, where their origin lies.

At $t\simeq 180~{\rm{sec}}$ ($T\simeq 1~{\rm{MeV}}$),
nucleosynthesis takes place, i.e. protons and neutrons form
nuclei. The abundance of light elements (D, $^{3}{\rm He}$,
$^{4}{\rm He}$, $^{6}{\rm Li}$, and $^{7}{\rm Li}$) depends 
(see e.g. Ref.~\cite{peebles}) crucially on the number of 
light particles (with mass $\lesssim
1~{\rm{MeV}}$), i.e. the number $N_{\nu}$ of light neutrinos
and the baryon abundance $\Omega_{\rm B}h^{2}$ ($\Omega_{\rm B}
=\rho_{\rm B}/\rho_{\rm c}$ with $\rho_{\rm B}$ being the baryon 
energy density). Agreement with 
observations \cite{deuterium} can be achieved for $N_{\nu}=3$ 
and $\Omega_{\rm B}h^{2}\simeq 0.02$. This is the third success 
of SBB cosmology. Much later, at the so-called equidensity 
time $t_{\rm{eq}}\simeq 4.7\times 10^4~{\rm{years}}$, 
matter dominates over radiation. At $t\simeq 200,000~h^{-1}~
{\rm{years}}$ ($T\simeq 3,000~{\rm{K}}$), the decoupling of 
matter and radiation and the recombination of atoms occur. 
After this, radiation evolves independently and is detected 
today as CMBR with temperature $T_{0}\simeq 2.73~{\rm{K}}$. 
The existence of the CMBR is the fourth success of SBB. 
Finally, structure formation \cite{structure} starts at 
$t\simeq 2\times 10^{8}~{\rm{years}}$.

\section{Shortcomings of SBB cosmology}
\label{sec:puzzles}

The SBB model has been very successful in explaining, among 
other things, the Hubble expansion of the universe, the 
existence of the CMBR, and the abundance of the light elements 
formed during nucleosynthesis. Moreover, combined with GUTs 
provides the basis for generating the baryon asymmetry of the 
universe, i.e. the slight excess of baryons over antibaryons,
so that after baryon-antibaryon annihilation there are leftover
baryons out of which the visible part of the universe is made.
Despite its successes, this model had a number of very puzzling
shortcomings which we will now summarize:

\begin{enumerate}

\item \textit{The horizon problem:} The CMBR was emitted at 
the decoupling of matter and radiation when $T_{\rm d}\simeq
3,000~\rm{K}$. The decoupling time $t_{\rm d}$ is 
estimated from
\begin{equation}
\frac{T_0}{T_{\rm d}}
=\frac{a(t_{\rm d})}{a(t_0)}=\left(\frac {t_{\rm d}}{t_0}
\right)^\frac{2}{3}\simeq\frac{2.73~\rm{K}}{3,000~\rm{K}}
\label{eq:dec}
\end{equation}
and turns out to be $t_{\rm d}\simeq 200,000~h^{-1}$ years.
The distance over which the CMBR has traveled since its 
emission is
\begin{equation}
a(t_0)\int^{t_{0}}_{t_{\rm d}}\frac{dt^\prime}
{a(t^\prime)}=3t_0\left[1-\left(\frac{t_{\rm d}}{t_0}
\right)^\frac{2}{3}\right]\simeq 3t_0\simeq
6,000~h^{-1}~\rm{Mpc},
\label{eq:lss}
\end{equation}
which practically coincides with $d_H(t_0)$. A sphere of 
radius $d_H(t_0)$ around us is called the {\it last scattering 
surface} since the CMBR
has been emitted from it. The particle horizon at decoupling
$3t_{\rm d}\simeq 0.168~h^{-1}~\rm{Mpc}$, expanded until now 
to become $0.168~h^{-1} (a(t_0)/a(t_{\rm d}))~{\rm{Mpc}}
\simeq 184~h^{-1}~\rm{Mpc}$. The angle subtended by this 
decoupling horizon now is $\vartheta_{\rm d}\simeq 184/6,000
\simeq 0.03~\rm{rads}$. Thus, the sky splits into 
$4\pi/(0.03)^2\simeq 14,000$ patches which never communicated 
causally before emitting the CMBR. The puzzle then is how 
can the temperature of the black body radiation from these 
patches be so finely tuned as the measurements of the 
Planck satellite \cite{planck} require.

\item \textit{Flatness Problem:} The present energy density of 
the universe has been observed \cite{planck} to be very close to 
its critical value corresponding to a flat universe ($\Omega_{0}=
1\pm 0.001$). From Eq.~(\ref{eq:friedmann}), we obtain 
$(\rho-\rho_{\rm c}) /\rho_{\rm c}=3(8\pi G\rho_{\rm c})^{-1}
(k/a^2)\propto a$ for a matter dominated universe. Thus, 
in the early universe, $|(\rho-\rho_{\rm c})/\rho_{\rm c}|
\ll 1$ and the question is why the initial energy density of the 
universe was so finely tuned to its critical value.

\item\textit{Magnetic Monopole Problem:} This problem arises 
only if we combine SBB with GUTs \cite{ggps} which predict the 
existence of magnetic monopoles \cite{monopole}. These monopoles  
are produced during the GUT phase transition, where the Higgs 
field $\phi$ responsible for the breaking of the GUT gauge 
symmetry $G$ acquires its VEV. They are localized deviations 
from the vacuum with radius $\sim M_{\rm G}^{-1}$ and mass 
$m_{\rm M}\sim M_{\rm G}/\alpha_{\rm G}$ ($\alpha_{\rm G}= 
g^2_{\rm G}/4\pi$, where $g_{\rm G}$ is the GUT gauge 
coupling constant). The value of $\phi$ on a sphere $S^2$ of 
radius $\gg M_{\rm G}^{-1}$ around the monopole lies in the vacuum 
manifold $G/G_{\rm SM}$ and we, thus, obtain a mapping: $S^2
\rightarrow G/G_{\rm SM}$. If this mapping is homotopically 
non-trivial, the monopole is topologically stable. The initial 
relative monopole number density must satisfy the
causality bound \cite{einhorn} $r_{\rm M,in}\equiv
(n_{\rm M}/T^3)_{\rm{in}}\gtrsim \rm{10^{-10}}$
($n_{\rm M}$ is the monopole number density), which comes from the
requirement that, at monopole production, $\phi$ cannot be
correlated at distances bigger than the particle horizon. The
subsequent evolution of monopoles is studied in
Ref.~\cite{preskill}. The result is that, if $r_{\rm M,in}
\gtrsim\rm{10^{-9}}$ ($\lesssim
\rm{10^{-9}}$), the final relative monopole number density
$r_{\rm M,fin}\sim 10^{-9}$ ($\sim r_{\rm M,in}$). This combined
with the causality bound yields $r_{\rm M,fin} \gtrsim 
\rm{10^{-10}}$. However, the requirement that monopoles
do not dominate the energy density of the universe at
nucleosynthesis gives
\begin{equation}
r_{\rm M} (T \simeq 1~\rm{MeV}) \lesssim \rm{10^{-19}}
\label{eq:nucleo}
\end{equation}
and we obtain a clear discrepancy of about nine orders of
magnitude.

\item\textit{Density Perturbations:} For structure formation 
\cite{structure} in the universe, we need a primordial density 
perturbation $\delta \rho/\rho$ at all length scales with a 
nearly flat spectrum \cite{hz}. We also need an explanation of 
the temperature fluctuations of the CMBR measured by the 
Planck satellite \cite{planck} at angles 
$\vartheta\gtrsim 
\vartheta_{\rm d}\simeq 2^{o}$, which violate causality. The 
SBB model cannot provide the required perturbations.

\end{enumerate}

\section{Inflation}
\label{sec:inflation}

All these four puzzles are solved by inflation 
\cite{guth,bookinflation}, which is a period of exponential
expansion in the early universe. Consider a real scalar 
field $\phi$ (the inflaton) with a (symmetric) potential 
$V(\phi)$ which is quite flat near $\phi=0$ and has minima at 
$\phi=\pm\langle\phi\rangle$ with $V(\pm\langle\phi\rangle)=0$. 
At high $T$'s, the potential acquires temperature corrections,
which make $\phi=0$ the absolute minimum. As the temperature 
drops, the effective potential tends to the zero-temperature 
potential, but a small barrier separating the local minimum at 
$\phi=0$ and the vacua at $\phi=\pm\langle\phi\rangle$ remains. 
At some point, $\phi$ tunnels out to $\phi_1\ll\langle\phi
\rangle$ and a bubble with $\phi=\phi_1$ is created in the 
universe. The field then rolls over to the minimum of $V(\phi)$ 
very slowly (due to the flatness of the potential $V(\phi)$) 
with the energy density $\rho\simeq V(\phi=0)\equiv V_0$ 
remaining practically constant for quite some time. The 
energy-momentum tensor during the slow roll-over of the 
inflaton becomes 
$T_{\mu}^{~\nu}\simeq -V_{0}~\delta_{\mu}^{~\nu}$ yielding 
$\rho\simeq -p\simeq V_0$. So, the pressure is practically 
opposite to the energy density, which remains almost constant 
in accord with Eq.~(\ref{eq:continuity}). 
The scale factor $a(t)$ grows (see below) and the curvature 
term $k/a^2$ in Eq.~(\ref{eq:friedmann}) diminishes. We 
thus obtain $H^2\simeq 8\pi GV_0/3={\rm constant}$, which gives 
$a(t)\propto e^{Ht}$. Therefore, the bubble expands 
exponentially for some time and $a(t)$ grows by a factor
\begin{equation}
\frac {a(t_{\rm f})}{a(t_{\rm i})}={\rm{exp}}
H(t_{\rm f}-t_{\rm i})
\equiv{\rm{exp}}H\tau
\label{eq:efold}
\end{equation}
between an initial ($t_{\rm i}$) and a final ($t_{\rm f}$) time.

It is almost obvious that, with an adequate number of 
$e$-foldings $N\equiv H\tau$, inflation automatically 
resolves the first three puzzles of SBB:   

\begin{enumerate}
 
\item\textit{Resolution of the Horizon Problem:} The 
particle horizon during inflation
\begin{equation}
d_H(t)=e^{Ht}\int^t_{t_{\rm i}} \frac{d
t^\prime}{e^{Ht^\prime}}\simeq H^{-1}{\rm{exp}}H(t-t_{\rm i}),
\label{eq:horizon}
\end{equation}
for $t-t_{\rm i}\gg H^{-1}$, grows as fast as $a(t)$. At $t_{\rm f}$
where inflation ends, $d_H(t_{\rm f})\simeq H^{-1}{\rm{exp}}
H\tau$ and $\phi$ starts oscillating about the vacuum. It 
then decays and reheats \cite{reheat} the universe at a 
temperature $T_{\rm r}\sim 10^9~{\rm{GeV}}$ \cite{gravitino} 
after which normal big bang cosmology is recovered. The 
particle horizon at the end of inflation $d_H(t_{\rm f})$ is 
stretched during the $\phi$-oscillations by a factor 
$\sim 10^9$ and between $T_{\rm r}$
and the present time by a factor $T_{\rm r}/T_0$. So, it finally
becomes equal to $H^{-1}e^{H\tau}10^9(T_{\rm r}/T_0)$, which must
exceed $2H_{0}^{-1}$ if the horizon problem is to be solved.
This readily holds for $V_0\sim M_{\rm G}^{4}$,
$M_{\rm G}\sim 10^{16}~{\rm GeV}$, and
$N\gtrsim 55$.

\item\textit{Resolution of the Flatness Problem:}
The curvature term of the Friedmann equation, at present, 
is
\begin{equation}
\left(\frac{k}{a^2}\right)_0\simeq\left(\frac{k}{a^2}
\right)_{\rm bi}e^{-2H\tau}~10^{-18}\left(\frac {10^{-13}~
{\rm{GeV}}}{10^9~{\rm{GeV}}}\right)^2,
\label{eq:curvature}
\end{equation}
where the factors in the RHS are the curvature term 
before inflation and its growth factors during inflation,
$\phi$-oscillations, and after reheating. Assuming
$(k/a^2)_{\rm bi}\sim H ^2$, we obtain $\Omega_0-1
=k/a_{0}^{2}H_{0}^{2}\sim 10^{48}~e^{-2H \tau}\ll 1$ for 
$H\tau\gg 55$. Strong inflation implies that the present 
universe is flat with a great accuracy.

\item\textit{Resolution of the Monopole Problem:} For 
$N\gtrsim 55$, the magnetic monopoles are
diluted by at least 70 orders of magnitude and become 
irrelevant. Also, since $T_{\rm r} \ll m_{\rm M}$, there 
is no magnetic 
monopole production after reheating. For models leading to a 
possibly measurable magnetic monopole density, see e.g.
Refs.~\cite{inter2,thermalinf,stringsmono,extended}.

\item\textit{Generation of density perturbations:} Inflation 
transforms the quantum fluctuations of the almost massless 
inflaton field into classical metric perturbations, which in 
turn generate tiny primordial density perturbations $\delta
\rho/\rho\simeq 5.6\times 10^{-5}$. These perturbations grow 
in the late universe to become non-linear and lead to the 
formation of structure (galaxies, filaments, and great voids) 
via gravitational collapse of matter. They also generate the 
temperature fluctuation $\delta T/T\simeq 6\times 10^{-6}$ in 
the CMBR measured by the COBE, WMAP, and Planck satellites. The 
predictions of inflation can fully agree with all experimental 
findings.

\end{enumerate} 

\section{Details of Inflation}
\label{sec:detail}

The Hubble parameter during inflation is slowly varying with 
the value of $\phi$:
\begin{equation}
H^{2}(\phi)=\frac{8\pi G}{3}V(\phi).
\label{eq:hubble}
\end{equation}
The evolution equation for $\phi$  is
\begin{equation}
\ddot{\phi}+3H\dot{\phi}+\Gamma_{\phi}\dot{\phi}+
V^{\prime}(\phi)=0,
\label{eq:evolution}
\end{equation}
where the dot and prime denote derivation with respect to 
the cosmic time and $\phi$, respectively, and $\Gamma_{\phi}$ 
is the decay width \cite{width} of the inflaton to light 
particles, which is assumed to be weak ($\Gamma_{\phi}\ll H$) 
\cite{dh}. Inflation is by definition the situation where the 
{\it friction term} $3H\dot{\phi}$ dominates and 
Eq.~(\ref{eq:evolution}) reduces to the inflationary equation 
\cite{slowroll}
\begin{equation}
3H\dot{\phi}=-V^{\prime}(\phi).
\label{eq:infeq}
\end{equation}
The conditions for this equation to hold (slow roll conditions)
are:
\begin{equation}
\epsilon,~|\eta|\leq 1~~{\rm with}~~\epsilon\equiv
\frac{1}{2}m_{\rm P}^{2}\left(\frac{V^{\prime}(\phi)}
{V(\phi)}\right)^{2},~\eta\equiv m_{\rm P}^{2}
\frac{V^{\prime\prime}(\phi)}{V(\phi)}.
\label{eq:src}
\end{equation}
The end of the slow roll-over occurs when either of these
inequalities is saturated. If $\phi_{\rm f}$ is the value of 
$\phi$ at the end of inflation, then $t_{\rm f} \sim H^{-1}
(\phi_{\rm f})$.

The number of $e$-foldings during inflation from $t_{\rm i}$ 
( or $\phi_{\rm i}$) to $t_{\rm f}$ ( or $\phi_{\rm f}$) is
\begin{equation}
N(\phi_{\rm i}\rightarrow \phi_{\rm f})\equiv\ln 
\left(\frac{a(t_{\rm f})}
{a(t_{\rm i})}\right)=\int^{t_{\rm f}} _{t_{\rm i}} Hdt=
\int^{\phi_{\rm f}}_{\phi_{\rm i}}\frac{H (\phi)}
{\dot{\phi}}d\phi=-\int^{\phi_{\rm f}}_{\phi_{\rm i}} \frac 
{3 H^2 (\phi)d \phi}{V^{\prime}(\phi)},
\label{eq:nefolds}
\end{equation}
where Eqs.~(\ref{eq:efold}) and (\ref{eq:infeq}) were used. 

After the end of inflation at $t_{\rm f}$, $\phi$ starts performing 
coherent damped oscillations about the global minimum of the 
potential. The rate of energy density loss, due to the 
expansion of the universe, is given by
\begin{equation}
\dot{\rho}=-3H\dot{\phi}^2= -3H(\rho+p)=-3H\gamma\rho,
\label{eq:damp}
\end{equation}
where $\rho=\dot{\phi}^2/2+V(\phi)$, $p=\dot{\phi}^2/2-
V(\phi)$, and we averaged $p$ over one oscillation of
$\phi$.  From this equation and Eq.~(\ref{eq:friedmann}), 
we obtain 
\begin{equation}
\rho\propto a^{-3\gamma},\quad a(t)\propto t^{\frac{2}
{3\gamma}}.
\end{equation} 
The parameter $\gamma$ depends on the shape of the potential 
$V(\phi)$. For a quadratic (quartic) potential, $\gamma=1$ 
($\gamma=4/3$) and the expansion is similar to that of a 
matter (radiation) dominated universe.

In order to discuss the subsequent decay of the inflaton, 
we must use its full evolution equation in 
Eq.~(\ref{eq:evolution}), which also includes the decay 
term $\Gamma_{\phi} \dot{\phi}$. This is approximately 
solved \cite{reheat,oscillation} by
\begin{equation}
\rho(t)=\rho_{\rm f} \left(\frac{t}{t_{\rm f}}\right)^{-2}
e^{-\Gamma_{\phi}t},
\label{eq:rho}
\end{equation}
where $\rho_{\rm f}$ is the energy density at $t_{\rm f}$ and, for 
definiteness, we considered that the potential is quadratic. 
The second and third factors in the RHS of this equation 
represent the dilution of the field energy due to the 
expansion of the universe and the decay of $\phi$ to light 
particles respectively.

All pre-existing radiation (known as old radiation) was 
diluted by inflation, so the only radiation present is the 
one produced by the decay of $\phi$ and is known as new 
radiation. Its energy density $\rho_{\rm r}$ satisfies 
\cite{reheat,oscillation} the equation
\begin{equation}
\dot{\rho}_{\rm r}=-4H \rho_{\rm r}+\gamma\Gamma_{\phi}\rho,
\label{eq:newrad}
\end{equation}
where the first term in the RHS represents the dilution of
radiation due to the cosmological expansion while the second 
one is the energy density transfer from $\phi$ to radiation. 
This yields
\begin{equation}
\rho_{\rm r}=\frac {3}{5}~\rho~\Gamma_{\phi}t\left[1+
\frac{3}{8}~\Gamma_{\phi}t+\frac
{9}{88}~(\Gamma_{\phi}t)^2+\cdots\right]
\label{eq:expand}
\end{equation}
with $\rho=\rho_{\rm f} (t/t_{\rm f})^{-2}{\rm{exp}} 
(-\Gamma_{\phi}t)$
being the energy density of the field $\phi$. Initially, 
$\rho_{\rm r}\ll \rho$, but, at $t_{\rm d}=
\Gamma_{\phi}^{-1}$, $\rho_{\rm r}$ 
dominates and the universe becomes radiation dominated. 
The temperature at $t_{\rm d}$ is
\begin{equation}
T_{\rm r}=\left(\frac {45}{2 \pi^{2}g_*}
\right)^\frac{1}{4}(\Gamma_{\phi}m_{\rm P})^\frac{1}{2}
\label{eq:reheat}
\end{equation}
and is historically called reheat temperature although 
there is neither supercooling nor reheating.

As already mentioned, inflation not only homogenizes the 
universe but also generates the density perturbations 
needed for structure formation. An important notion for 
understanding the underlying mechanism is event horizon. 
It includes 
all points with which we will eventually communicate 
sending signals at $t$. In contrast to the case of 
matter or radiation dominance, the instantaneous radius 
of the event horizon during inflation is finite:
\begin{equation}
d_{e}(t)=a(t)\int^{\infty}_{t}
\frac{dt^{\prime}}{a(t^{\prime})}=H^{-1}<\infty.
\label{eq:event}
\end{equation}
Points in our event horizon at $t$ are eventually
pulled out by the exponential expansion. We say that 
these points (and the corresponding scales) crossed 
outside the event horizon. Actually, the situation is 
like in a black hole turned inside out. Then, exactly 
as in a black hole, there are quantum fluctuations of 
the thermal type governed by the Hawking temperature
\cite{hawking} $T_{\rm H}=H/2\pi$. It turns out
\cite{bunch} that the quantum fluctuations of the
inflaton are $\delta\phi=T_{\rm H}$ leading to density 
perturbations $\delta\rho= V^{\prime}(\phi)\delta\phi$. 
As the scale $\ell$ of these perturbations crosses 
outside the event horizon, they become \cite{fischler} 
classical. When $\ell$ (or $k=2\pi/\ell$) re-enters the
post-inflationary horizon, we obtain the perturbation
\cite{zeta} (for a review, see e.g. Ref.~\cite{zetarev})
\begin{equation}
\left(\frac{\delta\rho}{\rho}\right)_{\ell}=
\frac{1}{5\sqrt{3}\pi}~\frac {V^\frac{3}{2}
(\phi_{\ell})}{m^{3}_{\rm P} V^{\prime}(\phi_{\ell})},
\label{eq:deltarho}
\end{equation}
where $\phi_{\ell}$ is the value of $\phi$ when $\ell$
crosses outside the horizon. From the results of COBE 
\cite{cobe}, $(\delta\rho/\rho)\simeq 2\times 10^{-5}$
at the pivot scale $k=0.002~{\rm Mpc}^{-1}$. The spectrum
of the perturbations $\delta\rho/\rho\propto\ell^{\alpha_{\rm s}}$
is characterized by the scalar spectral index 
$n_{\rm s}=1-2\alpha_{\rm s}=1+2\eta-6\epsilon$. The 
Harrison-Zeldovich 
flat spectrum \cite{hz} corresponds to $n_{\rm s}=1$, while the 
experimental value is $n_{\rm s}=0.968\pm 0.006$ \cite{planck}.

The density fluctuations on the last scattering surface 
produce temperature fluctuations in the CMBR. The dominant 
mechanism is the scalar Sachs-Wolfe \cite{sachswolfe}
effect: regions with a deep gravitational potential will 
cause the photons to lose energy as they climb up the 
potential well and, thus, appear cooler. The scalar 
quadrupole anisotropy is
\begin{equation}
\left(\frac{\delta T}{T}\right)_{\rm S}=\frac{1}
{12\sqrt{5}\pi}
\frac{V^\frac{3}{2}(\phi_{\ell})}{m^{3}_{\rm P}
V^{\prime}(\phi_{\ell})}.
\label{eq:quadrupole}
\end{equation}
There are also tensor fluctuations \cite{tensor} in the
temperature of the CMBR with the tensor quadrupole 
anisotropy given by
\begin{equation}
\left(\frac{\delta T}{T}\right)_{\rm T}
\simeq 0.15~\frac
{V^{\frac{1}{2}}(\phi_{\ell})}{m^{2}_{\rm P}}.
\label{eq:tensor}
\end{equation}
The tensor-to-scalar ratio
\begin{equation}
r=\frac{\left(\delta T/T\right)^{2}_{T}}
{\left(\delta T/T\right)^{2}_{S}}
\label{eq:ratio}
\end{equation}
is experimentally bound to be $\lesssim 0.06$ 
\cite{planck}. 

\section{Baryogenesis}
\label{sec:baryo}

A successful inflationary scenario has to be followed by 
a successful reheating process \cite{reheat}. Note that, 
in the case of supersymmetric theories, the reheat 
temperature $T_{\rm r}\lesssim 10^{9}~{\rm GeV}$ since 
otherwise the gravitino relic density is unacceptably 
large \cite{gravitino}. Also the observed \cite{planck} 
baryon asymmetry $n_{\rm B}/s\simeq 8.66\times 10^{-11}$ 
\cite{planck} should be generated after inflation ($n_{\rm B}$ 
and $s$ are the baryon number and entropy densities). 
The most promising way to generate the baryon asymmetry 
after inflation is via non-thermal 
\cite{nonthermalepto,vlachos} leptogenesis 
\cite{thermallepto} (for a review, see 
Ref.~\cite{pedestrians}). In accordance with this scenario,
the inflaton decays into a pair of right handed neutrinos 
($\nu^c$). The subsequent out-of-equilibrium decay of these 
$\nu^c$'s into (anti)leptons and electroweak Higgs fields can 
generate a primordial lepton asymmetry provided there is 
CP-violation. This asymmetry, at the electroweak transition, is 
partly turned \cite{dimopoulos} into baryon asymmetry by 
sphaleron \cite{sphaleron} effects.

In order to understand the sphaleron mechanism, let us 
recall that the vacua of the $SU(2)_{\rm L}$ gauge theory 
are characterized \cite{vacuum} by a winding number 
$n\in Z$. In the presence of the VEV of the electroweak 
doublet which breaks $SU(2)_{\rm L}$, the minimal height 
of the potential barrier between the vacua with
winding numbers $n$ and $n+1$ corresponds to a static
solution \cite{sphaleronsol} called {\it sphaleron}, which 
is actually a saddle point of the potential. The mass of 
the sphaleron is about $10-15~{\rm TeV}$. At $T=0$, the 
tunneling between the vacua $|n\rangle$ and $|n+1\rangle$
is utterly suppressed. However, between $T\sim 200~
{\rm GeV}$ and the critical temperature $T_{\rm c}$ of the 
electroweak phase transition, this tunneling is very 
frequent and in equilibrium with the universe 
\cite{dimopoulos,sphaleron}. Baryon ($B$) and Lepton 
($L$) number anomalies imply \cite{thooft} that the 
tunneling from $|n\rangle$ to $|n+1\rangle$ is accompanied
by a change $\Delta B=\Delta L=-3$. Note that $B-L$ is not
violated and the primordial asymmetry in $-L$ can be 
considered as primordial $B-L$ asymmetry, which remains
constant at subsequent times. At the electroweak phase
transition this asymmetry is partly converted into $B$ 
asymmetry. The resulting $n_{\rm B}/s$ can be calculated
\cite{turneribanez} by setting equal to zero the algebraic 
sum of the chemical potentials ($\mu$) of all the particles 
involved in each of the reactions which are in thermal 
equilibrium. Solving the system of the resulting equations,
we find the fraction of the primordial $B-L$ asymmetry which
turns into $B$ asymmetry.

\section{Dark Energy}
\label{sec:DE}

Inflation implies that the present universe is exactly flat 
($\Omega=\rho/\rho_{\rm c}= 1$) and thus will keep expanding for 
ever. This is also confirmed by the recent measurements 
\cite{planck}, which also reveal that matter is only $27\%$ 
of the universe. The question then arises what is the rest 
$73\%$ of the universe made of. The answer to this question 
was first given in 1997-8 by the observations on 
Supernovae Ia \cite{supernovae}, which are seen today as they 
were in a previous cosmic time. The unexpected result of these 
observations was that the expansion of the universe in previous 
times was slower than it is today, i.e. the expansion of 
the universe is accelerating. This can be explained if more 
than 2/3 of the energy in the universe is in the form of 
cosmological constant, i.e. it is practically not diluted by the 
expansion exactly as the energy driving inflation (because of 
the existence of negative pressure close in magnitude to the 
energy density). One can then wonder whether we are about to 
enter a new inflationary phase. The idea of quintessence 
\cite{quintessence} (for a review, see Ref.~\cite{quintreview})
though tells us that this may not be the case. The WMAP and 
Planck satellites \cite{wmap,planck} confirm that the $73\%$ of 
the energy density in the universe is in the form of dark energy, 
i.e. a form close to a cosmological constant.

\section{Dark Matter}
\label{sec:DM}

Studies of nucleosynthesis \cite{bbn} combined with the results 
of the Planck satellite \cite{planck} imply that the baryonic
(visible) matter constitutes only $4.85\%$ of the energy density 
in the universe. Therefore, if one subtracts baryonic matter 
from the total matter content of the universe, which is about 
$27\%$, concludes that about $22\%$ of the universe consists of 
dark matter. The question then arises what is the nature of dark 
matter. Dark matter is usually considered to be cold, i.e. 
consisting of non-relativistic massive particles which interact 
very weakly (mainly gravitationally) with all other particles. 
However, very light weakly interacting particles such as axions 
can also contribute to dark matter since they are non-relativistic 
for different reasons. Also warm or hot dark matter has been 
considered.   

What could the dark matter particle be? There are several
proposal, but I will mention only the most important ones.

\begin{enumerate}

\item\textit{Lightest supersymmetric particle (LSP):} This 
particle is the most promising candidate for dark matter
\cite{LSP} since it is protected by a discrete $Z_2$ R-parity 
symmetry from decaying into lighter particles. The LSPs can 
annihilate in pairs or coannihilate with the next-to-LSPs. 
There are many detailed calculations of the relic density of 
the LSPs (see e.g. Refs.~\cite{LSPrelic,micromegas}) which 
show that, under certain circumstances, they can account for 
the dark matter. 

\item\textit{Axion:} This light boson is connected with the 
Peccei-Quinn solution \cite{PecceiQuinn} of the strong CP 
problem and can contribute \cite{axionDM} to dark matter. 
However, one has to be careful since, if combined with 
inflation, it may lead to unacceptably large isocurvature 
perturbations \cite{iso} in the CMBR. This problem is avoided 
if the Peccei-Quinn symmetry breaking occurs after the end of 
inflation \cite{interDM} or if the value of the Peccei-Quinn 
field during inflation is kept large \cite{axionmuterm}.

\item\textit{Superheavy fermions:} Intermediate scale dark 
matter particles coexisting with axions appear in 
Refs.~\cite{interDM,inter2}, while superheavy dark matter  
particles (wimpzillas) have been studied in 
Ref.~\cite{wimpzilla}.

\end{enumerate} 

The LSPs could be detected by their scattering off nuclei 
as in the XENON 1T experiment \cite{XENON}, which has no 
positive results so far. The detection of axions is even
harder. In supersymmetric theories the gravitino 
\cite{gravitinoDM}, the superpartner of graviton, or the 
axino \cite{axino} can also contribute to the composition 
of dark matter. 

The most promising candidate for dark matter is the LSP. 
In supersymmetric theories, there exist a discrete $Z_2$ 
symmetry known as R-parity, which is necessary for 
preventing unacceptably fast proton decay. Under R-parity, 
all the standard model particles are even, while all the 
supersymmetric particles are odd. By virtue of R-parity, 
the LSP, which is normally the lightest neutralino, is 
stable as it cannot decay to lighter particles and thus 
is a good candidate for dark matter. The supersymmetric 
particles can annihilate into pairs to standard model 
particles. To find the LSP relic abundance, we need to 
consider also coannihilations of the LSP with the 
next-to-LSP, which is usually the lightest stau. This is 
a very complicated 
calculation since very many processes are involved but,
fortunately, there are already publicly available codes 
for this calculation such as microMEGAs \cite{micromegas}.
The relic density of neutralinos usually comes out too
large. However, the following main mechanisms can
reduce the LSP relic abundance to the observed value of
dark matter abundance $\Omega_{\rm DM}\,h^2\simeq 0.12$ 
\cite{planck}.

\begin{enumerate}

\item\textit{$A$-pole exchange:} The LSP pair annihilation can 
be resonantly enhanced by an $A$-pole exchange in the 
$s$-channel \cite{Apole} ($A$ is the CP-odd Higgs boson in the 
minimal supersymmetric standard model).

\item\textit{Coannihilation:} Coannihilation \cite{coannih} 
of the LSP with 
the next-to-LSP, which is usually the lightest stau eigenstate,
can reduce drastically the relic LSP density. This, however, 
requires a small mass splitting between the LSP and the 
next-to-LSP.

\end{enumerate} 

\section{Quantum Gravity} 
\label{sec:qg}

So far, we considered gravity at the classical (non-quantum) 
level and separately from the other three interactions, which 
are described by relativistic quantum field theory and 
possibly unified in a GUT. The questions then arise:
\begin{list} 
\setlength{\rightmargin=0cm}{\leftmargin=0cm}
\item[{\bf (a)}] Can gravity be quantized as the other 
three interactions?
\item[{\bf (b)}] Can it be unified with the other three 
interactions to yield {\it the theory of everything}?
\end{list}
Quantum gravity phenomena are expected to appear at very 
high energies of the order of the Planck mass $m_{\rm P}
\simeq 2.44\times 10^{18}~{\rm GeV}$, or at small 
distances of the order of the Planck length $\ell_{\rm P}
\equiv m_{\rm P}^{-1}\sim 10^{-33}~{\rm cm}$, or at small 
times of the order of $t_{\rm P}\sim 10^{-43}~{\rm sec}$. 
Therefore, quantum fluctuations of gravity are expected 
to be important
\begin{list} 
\setlength{\rightmargin=0cm}{\leftmargin=0cm}
\item[{\bf (i)}] very near the big bang, i.e. for cosmic 
times $t \lesssim t_{\rm P}$, and
\item[{\bf (ii)}] for the possible generation of primordial 
gravity waves, which could originate from the quantum 
fluctuations of gravity during cosmological inflation. These 
waves may be detected in the future by the space based 
observatory LISA.
\end{list}

In trying to construct a quantum field theory for gravity, 
we encounter problems with {\it renormalizability}. In order 
to understand the meaning of renormalizability, let us consider 
classical electrodynamics, as an example. Relativity implies 
that every elementary particle must be point-like. Consequently, 
the energy of the electric field ${\bf\bar{E}}$ around such a 
particle with charge $e$ (e.g. the electron) is proportional to 
\begin{equation}
\int_0^\infty {\bf\bar{E}}^2 4\pi r^2 dr=
4\pi e^2\int_0^\infty\frac{dr}{r^2}.
\end{equation}
This integral diverges at $r=0$ and thus the energy is infinite.
We must then assume that the {\it bare} mass of the particle is 
$-\infty$, so that the final mass is $\infty-\infty$ and turns 
out to be finite.

This situation appears also in quantum electrodynamics, where 
we encounter several meaningless infinities while performing 
various calculations. However, all these infinities can be 
systematically gathered and put inside two parameters: the mass 
and charge of the electron. We then take the bare values of 
these parameters to be infinite, so that their final 
{\it renormalized} values are finite and equal to their 
experimental values. This complicated mathematical procedure is 
called {\it renormalization} and the theories where it applies 
are called renormalizable. The three interactions (except 
gravity) are renormalizable as shown by 't Hooft and Veltman 
(for a review, see Ref.~\cite{renorm}). Therefore, with a finite 
number of experiments, we can determine the 
values of a finite number of parameters and then everything else 
can be predicted. So, the theory has a high predictability. The 
problem with quantum gravity is that, as it turns out, it is not 
renormalizable.

\section{String Theory}
\label{sec:string}

The theory of (super)strings \cite{superstrings} was proposed 
in order to cure the non-renormalizabity of gravity and also 
unify it with the other three interactions, so as to achieve 
the construction of the theory of everything.
The idea is that the fundamental objects are not point-like 
particles, but one dimensional strings. Adding  supersymmetry to 
the scheme one obtains superstrings. We have strings of size   
$\ell_{\rm P}\sim 10^{-33}~{\rm cm}$ which vibrate, but 
for the present energies look like point particles. The various 
vibrational modes of these strings appear as particles with 
different quantum numbers and we obtain a unified description 
of all particles and all their interactions. The abandonment of 
point particles removes the notorious infinities. It is actually 
believed that string theory gives renormalizable or even 
completely finite unification of all four interactions (see e.g. 
Ref.~\cite{finite}). One of the vibrational modes is the graviton, 
which is the carrier of gravitational interactions. Therefore, we 
can, in principle, describe quantum gravity and the other 
three interactions in a unified way, which would be the theory of 
everything.

However, there are some important shortcomings of string theory.
\begin{list}
\setlength{\rightmargin=0cm}{\leftmargin=0cm}
\item[{\bf (i)}] Initially, people thought that there is a 
unique solution of string theory, which would make it of very 
high predictability. However, we now know that there exists a 
huge number of solutions ($\sim 10^{500}$). The string vacua 
comprise a collective {\it landscape} \cite{landscape} 
with a huge number of hills, valleys, etc and, therefore, a 
huge number of  
minima (solutions). One can then employ the so called 
{\it anthropic principle}, which states that the minimum 
(solution) corresponding to our universe had the right 
conditions so as to produce us eventually. In my opinion, 
this is not an acceptable scientific way of thinking. 
\item[{\bf (ii)}] Although there exist very many solutions
with a variety of discrete symmetries (see e.g. 
Ref.~\cite{classification}), none of them reproduces exactly 
our universe.
\end{list}

An important prediction of string theory is that there exist 
ten spacetime dimensions \cite{superstrings} (or eleven in the 
case of M theory \cite{mtheory}).
Six of them are compactified, i.e. they are strongly curved 
to form a 6-dimensional compact manifold of size $\ell_{\rm P}
\sim 10^{-33}~{\rm cm}$. The other four dimensions remain open 
and correspond to the usual spacetime dimensions. The 
geometric structure of the compactified dimensions determines 
many of the phenomena encountered in  the 4-dimensional 
spacetime.

As already mentioned strings allow us to discuss quantum 
gravity and thus give us the possibility to approach the
big bang, i.e. the initial singularity at $t=0$. Therefore,
we can consider cosmic times $t\lesssim t_{\rm P}\sim 
10^{-43}~{\rm sec}$  or even negative times before the big 
bang. One problem we can address, in the light of string 
theory, is the problem of initial conditions \cite{initial} 
for inflation.

Conventional inflation takes place at the GUT phase transition 
at cosmic time $t\sim 10^{-37}~{\rm sec} \gg t_{\rm P}$, where 
we need, as initial condition for inflation to start, a large 
region of size few times $H^{-1}$ within which the fields are 
homogenized taking appropriate values \cite{initsuper}. 
However, this region consists of many regions of smaller size 
$\ell$, which resulted from the expansion of regions of size 
$\ell_{\rm P}$ at the Planck time,
where causal communication could not exceed the distance 
$\ell_{\rm P}$. Therefore, a region of size $H^{-1}$ at the 
onset of inflation cannot be homogenized. This is again a 
problem of initial conditions. Its resolution requires 
a first stage of inflation near or before the big bang to 
provide the necessary homogenization for the onset of 
conventional inflation. One possible solution could be 
the pre-big-bang scenario \cite{preBB}. Imagine that we travel 
backwards in 
time and pass through the initial singularity at $t=0$ , which 
is though smoothed by strings, and enter the realm of negative 
times (before the big bang). There, during the motion of the 
universe towards the initial singularity, we have conditions of 
very high curvature and the extra dimensions contract and 
compactify. This causes inflation (accelerated expansion) in 
the four open dimensions. This can be a first stage of 
inflation providing the initial conditions for the conventional 
inflation or can be the main inflation producing the density 
perturbations too. We should be though very careful with these 
stringy considerations since the physics of strings is not fully 
understood or solved yet. Note that the problem of initial 
conditions for inflation can be solved by a conventional two 
stage inflation \cite{2stage} too.

Another application of quantum gravity could be the explanation 
of the origin of primordial gravity waves if such waves are 
detected. The BICEP2 experiment \cite{bicep2} observed the 
polarization vector pattern of the CMBR, which can be split 
into an $\bf\bar{E}$-mode resembling an electrostatic field
with sources and sinks and a $\bf\bar{B}$-mode resembling
a magnetostatic field with vortices. Subtracting the 
$\bf\bar{B}$-mode from polarized dust from the observed 
$\bf\bar{B}$-mode in CMBR, they found that there is a 
remaining $\bf\bar{B}$-mode, which could be due to primordial 
gravity waves from inflation. The subsequent Planck satellite 
measurements \cite{planck}, however, did not confirm this 
conclusion, which as it turned out underestimated the 
foreground from polarized dust. However, possible existence 
of primordial gravity waves, which may be measurable in the 
foreseeable future, cannot be excluded since the Planck 
measurements \cite{planck} put only a moderate upper bound on 
the tensor-to-scalar ratio $r\lesssim 0.06$. These waves could 
originate from tensor (gravitational) quantum fluctuations 
during inflation which become classical fluctuation (i.e. 
gravity waves) as they exit the  inflationary horizon. 
Therefore, quantum gravity may be required to understand the
origin of these waves.

\section{Conclusions}
\label{sec:concl}

We presented the SBB cosmological model together with its 
successes and shortcomings, which are resolved by inflation, 
i.e. a period of exponential expansion in the early stages of 
the universe evolution. This may have happened during the GUT 
phase transition at which the relevant Higgs field was 
displaced from its vacuum value. This field (inflaton) could 
then, for some time, roll slowly towards the vacuum providing 
an almost constant false vacuum energy density driving the 
exponential expansion. Inflation provides the density 
perturbations required for the generation of the large scale 
structure in the universe and the temperature fluctuations in 
the CMBR. After the end of inflation, the inflaton performs 
damped oscillations about the vacuum, decays, and reheats the 
universe.

We described how the observed baryon asymmetry of the universe 
is generated, in inflationary models, via non-thermal 
leptogenesis. The inflaton normally decays into right handed
neutrinos, whose subsequent out-of-equilibrium decay into 
leptons (antileptons) and electroweak Higgs fields produces 
a primordial lepton asymmetry. This asymmetry is then partly 
converted into baryon asymmetry by the electroweak sphaleron 
effects. Recent measurements revealed that the present universe 
is flat with matter constituting only $27\%$ of its energy 
density. The rest $73\%$ of the universe is in the form of 
dark energy. On the other hand, the baryonic (visible) matter 
constitutes only $4.85\%$ of the universe. The rest $22\%$ 
is made of dark matter. The most promising candidate for dark 
matter is the LSP since it is 
protected by a discrete $Z_2$ R-parity symmetry from decaying. 
These particles can annihilate in pairs or coannihilate with 
the next-to-LSPs and their relic density can be reduced to the 
observed value of dark matter abundance mainly by resonant pair 
annihilation or coannihilation if they are almost degenerate 
with the next-to-LSPs. Other possible dark matter candidates 
include the axion and superheavy or intermediate scale 
fermions.

We sketched briefly the renormalization problem encountered 
in trying to quantize gravity and its possible resolution by
string theory, whose aim is not only to construct a viable 
theory for quantum gravity but also to unify it with the 
other three interactions. The main disadvantage of string 
theory is that it admits a huge number of solutions, but 
none of them seems to reproduce exactly our universe. It 
predicts that there exist ten spacetime dimensions (or 
eleven in 
the case of M theory). Six of them are compactified, while 
the other four remain open and are the usual spacetime 
dimensions. The geometric structure of the compactified 
dimensions determines many of the phenomena in the 
4-dimensional spacetime.

String theory allows us to discuss physics very near or
even before the big bang, where the quantum fluctuations 
of gravity are present. Therefore, it can help in resolving 
the problem of initial conditions for inflation. In order 
to have inflation started, one needs a large homogeneous 
region around the time of the GUT phase transition. 
However, this region consists of many smaller regions 
which originate from homogenized regions around the Planck 
era, where the causal communication could not extend to 
distances larger than the Planck length. Therefore, a first 
stage of inflation near or before the big bang which 
provides the required homogenization at the onset of 
inflation is needed. Such a primordial inflation can take 
place in the pre-big-bang period. During the motion of the 
universe towards the initial singularity, we have conditions 
of very high curvature and the extra dimensions contract 
and compactify. This causes inflation in the four open 
dimensions. Another application of quantum gravity could be 
the explanation of the origin of primordial gravity waves 
from inflation if such waves are detected in the future, 
which is not excluded by the recent data. 

\def\plb#1#2#3{{\it Phys. Lett.} B
{\bf #1}~(#3)~#2}
\def\apjl#1#2#3{{\it Astrophys. J. Lett.}
{\bf #1}~(#3)~#2}
\def\apj#1#2#3{{\it Astrophys. J.}
{\bf #1}~(#3)~#2}
\def\jetpl#1#2#3{{\it JETP Lett.}
{\bf #1}~(#3)~#2}
\def\jetpsp#1#2#3{{\it JETP (Sov. Phys.)}
{\bf #1}~(#3)~#2}
\def\spss#1#2#3{{\it Sov. Phys. -Solid State}
{\bf #1}~(#3)~#2}
\def\jpa#1#2#3{{\it J. Phys.} A
{\bf #1}~(#3)~#2}
\def\pr#1#2#3{{\it Phys. Reports}
{\bf #1}~(#3)~#2}
\def\mnras#1#2#3{{\it Mon. Not. Roy. Astr. Soc.}
{\bf #1}~(#3)~#2}
\def\n#1#2#3{{\it Nature}
{\bf #1}~(#3)~#2}
\def\cmp#1#2#3{{\it Commun. Math. Phys.}
{\bf #1}~(#3)~#2}
\def\prsla#1#2#3{{\it Proc. Roy. Soc. London} A
{\bf #1}~(#3)~#2}
\def\ptp#1#2#3{{\it Prog. Theor. Phys.}
{\bf #1}~(#3)~#2}
\def\prd#1#2#3{{\it Phys. Rev.} D
{\bf #1}~(#3)~#2}
\def\prl#1#2#3{{\it Phys. Rev. Lett.}
{\bf #1}~(#3)~#2}
\def\npb#1#2#3{{\it Nucl. Phys.} B
{\bf #1}~(#3)~#2}
\def\jhep#1#2#3{{\it J. High Energy Phys.}
{\bf #1}~(#3)~#2}
\def\anj#1#2#3{{\it Astron. J.}
{\bf #1}~(#3)~#2}
\def\baas#1#2#3{{\it Bull. Am. Astron. Soc.}
{\bf #1}~(#3)~#2}
\def\grg#1#2#3{{\it Gen. Rel. Grav.}
{\bf #1~(#3)}~#2}
\def\stmp#1#2#3{{\it Springer Trac. Mod. Phys.}
{\bf #1}~(#3)~#2}
\def\ijmpa#1#2#3{{\it Int. J. Mod. Phys.} A
{\bf #1}~(#3)~#2}
\def\lnp#1#2#3{{\it Lect. Notes Phys.}
{\bf #1}~(#3)~#2}
\def\apjs#1#2#3{{\it Astrophys. J. Suppl.}
{\bf #1}~(#3)~#2}
\def\jcap#1#2#3{{\it J. Cosmol. Astropart. Phys.}
{\bf #1}~(#3)~#2}
\def\ijmpe#1#2#3{{\it Int. J. Mod. Phys.} E
{\bf #1}~(#3)~#2}
\def\aap#1#2#3{{\it Astron. Astrophys.}
{\bf #1}~(#3)~#2}
\def\epjc#1#2#3{{\it Eur. Phys. J.} C
{\bf #1}~(#3)~#2}


\begin{thebibliography}{99}

\bibitem{bbn} G.~Steigman, {\it Primordial 
nucleosynthesis: successes and challenges}, 
\ijmpe{15}{1}{2006}.

\bibitem{wkt}
S.~Weinberg, {\it Gravitation and Cosmology}, 
J.~Wiley and Sons, New York 1972; 
E.W.~Kolb and M.S.~Turner, {\it The Early Universe},
{\it Front. Phys.} {\bf 69} (1990) 1.

\bibitem{ggps} J.C.~Pati and A.~Salam, {\it  	
Is Baryon Number Conserved?}, \prl{31}{661}{1973};
H.~Georgi and S.~Glashow, {\it Unity of All Elementary 
Particle Forces}, \prl{32}{438}{1974}.

\bibitem{cobe} 
C.L.~Bennett et al., {\it 4-Year COBE DMR Cosmic 
Microwave Background Observations: Maps and Basic 
Results}, {\it Astrophys. J.} {\bf 464} (1996) L1.

\bibitem{wmap} 
G.~Hinshaw et al., {\it Nine-Year Wilkinson Microwave 
Anisotropy Probe (WMAP) Observations: Cosmological 
Parameter Results}, {\it Astrophys. J. Suppl.} {\bf 208} 
(2013) 19.

\bibitem{planck}
N.~Aghanim et al., {\it Planck 2018 results. VI. 
Cosmological parameters}, arXiv:1807.06209 [astro-ph.CO].

\bibitem{monopole} G.'t~Hooft, {\it Magnetic Monopoles 
in Unified Gauge Theories}, \npb{79}{276}{1974};
A.~Polyakov, {\it Particle Spectrum in the Quantum 
Field Theory}, \jetpl{20}{194}{1974}.

\bibitem{preskill} J.P.~Preskill, {\it Cosmological 
Production of Superheavy Magnetic Monopoles}, 
\prl{43}{1365}{1979}.

\bibitem{structure} P.J.E.~Peebles, {\it The Large-Scale
Structure of the Universe}, Princeton University Press, 
Princeton 1980;
G.~Efstathiou, {\it The Physics of the Early Universe},
in proceedings of the {\it 36th Scottish Universities Summer 
School in Physics}, Editors J.A.~Peacock et al., 
Adam-Higler, Bristol 1990.

\bibitem{guth} 
A.H.~Guth, {\it The Inflationary Universe: A Possible 
Solution to the Horizon and Flatness Problems}, 
\prd{23}{347}{1981}.

\bibitem{bookinflation} 
G.~Lazarides, {\it Introduction to cosmology}, in proceedings 
of {\it Corfu Summer Institute on Elementary Particle Physics, 
1998}\, PoS(corfu98)014 [hep-ph/9904502];
A.R.~Liddle and D.H.~Lyth, {\it 
Cosmological Inflation and Large-Scale Structure}, Cambridge
University Press, Cambridge 2000;
G.S.~Watson, {\it An exposition on inflationary cosmology},
astro-ph/0005003;
G.~Lazarides, {\it Inflationary cosmology}, 
\lnp{592}{351}{2002} [hep-ph/0111328];
G.~Lazarides, {\it Introduction to inflationary cosmology},
in proceedings of {\it Corfu Summer Institute on Elementary 
Particle Physics, 2001} [hep-ph/0204294];
V.A.~Rubakov, {\it Introduction to Cosmology}, in proceedings of 
{\it RTN Winter School on Strings,Supergravity and Gauge Theories, 
SISSA 2005} PoS(RTN2005)003;
A.R.~Liddle and D.H.~Lyth, 	
{\it The primordial density perturbation: Cosmology, 
inflation and the origin of structure}, Cambridge
University Press, Cambridge 2009. 

\bibitem{reheat} R.J.~Scherrer and M.S.~Turner, 	
{\it Decaying Particles Do Not Heat Up the Universe}, 
\prd{31}{681}{1985}.

\bibitem{dilution} G.~Lazarides, C.~Panagiotakopoulos, and 
Q.~Shafi, {\it Relaxing the Cosmological Bound on Axions},
\plb{192}{323}{1987};
G.~Lazarides, R.K.~Schaefer, D.~Seckel, and Q.~Shafi, {\it
Dilution of Cosmological Axions by Entropy Production},
\npb{346}{193}{1990}.

\bibitem{gravitino} M.Yu.~Khlopov and A.D.~Linde,
{\it Is it Easy to Save the Gravitino?}, 
\plb{138}{265}{1984};
J.R.~Ellis, J.E~Kim, and D.V.~Nanopoulos,
{\it Cosmological Gravitino Regeneration and Decay},
\plb{145}{181}{1984};
J.R.~Ellis, D.V.~Nanopoulos, and S.~Sarkar,
{\it The Cosmology of Decaying Gravitinos},
\npb{259}{175}{1985};
J.R.~Ellis, G.B.~Gelmini, J.L.~L\'{o}pez,
D.V.~Nanopoulos, and S.~Sarkar, {\it Astrophysical 
constraints on massive unstable neutral relic particles},
\npb{373}{399}{1992};
M.~Kawasaki and T.~Moroi, {\it Gravitino production in the 
inflationary universe and the effects on big bang 
nucleosynthesis}, \ptp{93}{879}{1995}.

\bibitem{nonthermalepto}
G.~Lazarides and Q.~Shafi, {\it Origin of matter in the 
inflationary cosmology}, {\it Phys. Lett.} B {\bf 258} 
(1991) 305;
G.~Lazarides, R.K.~Schaefer, and Q.~Shafi, {\it 
Supersymmetric inflation with constraints on superheavy 
neutrino masses},
{\it Phys. Rev.} D {\bf 56} (1997) 1324.

\bibitem{thermallepto} 
M.~Fukugita and T.T.~Yanagida, {\it Baryogenesis Without 
Grand Unification}, {\it Phys. Lett.} B {\bf 174} (1986) 45.

\bibitem{dimopoulos} S.~Dimopoulos and L.~Susskind, 	
{\it On the Baryon Number of the Universe},
\prd{18}{4500}{1978}.

\bibitem{sphaleron} V.A.~Kuzmin, V.A.~Rubakov, and
M.~Shaposhnikov, {\it On the Anomalous Electroweak Baryon 
Number Nonconservation in the Early Universe}, 
\plb{155}{36}{1985};
P.~Arnold and L.~McLerran, {\it Sphalerons, Small Fluctuations 
and Baryon Number Violation in Electroweak Theory}, 
\prd{36}{581}{1987}.

\bibitem{LSP}
H.~Goldberg, {\it Constraint on the Photino Mass from 
Cosmology}, {\it Phys. Rev. Lett.} {\bf 50} (1983) 1419;
J.R.~Ellis, J.S.~Hagelin, D.V.~Nanopoulos, K.A.~Olive, 
and M.~Srednicki, {\it Supersymmetric Relics from the Big 
Bang}, {\it Nucl. Phys.} B {\bf 238} (1984) 453.

\bibitem{Apole}
V.~Barger and C.~Kao, {\it Relic density of neutralino 
dark matter in supergravity models}, {\it Phys. Rev.} D 
{\bf 57} (1998) 3131;
J.R.~Ellis and K.A.~Olive, {\it Revisiting the Higgs Mass and 
Dark Matter in the CMSSM}, {\it Eur. Phys. J.} C {\bf 72} 
(2012) 2005.

\bibitem{coannih}
K.~Griest and D.~Seckel, {\it Three exceptions in the 
calculation of relic abundances}, {\it Phys. Rev.} D 
{\bf 43} (1991) 3191.

\bibitem{PecceiQuinn}
R.D.~Peccei and H.R.~Quinn, {\it CP Conservation in the 
Presence of Instantons}, {\it Phys. Rev. Lett.} {\bf 38} 
(1977) 1440;
S.~Weinberg, {\it A New Light Boson?}, {\it Phys. Rev. 
Lett.} {\bf 40} (1978) 223; 
F.~Wilczek, {\it Problem of Strong P and T Invariance in 
the Presence of Instantons}, {\it Phys. Rev. Lett.} {\bf 40} 
(1978) 279.

\bibitem{wimpzilla}
E.W.~Kolb, D.J.H.~Chung, and A.~Riotto, 
{\it WIMPzillas!}, {\it AIP Conf. Proc.} {\bf 484} (1999) 91 
[hep-ph/9810361].

\bibitem{interDM}  
G.~Lazarides and Q.~Shafi, {\it Axion Model with 
Intermediate Scale Fermionic Dark Matter}, arXiv:
2004.11560 [hep-ph].

\bibitem{inter2}
G.~Lazarides and Q.~Shafi, {\it Monopoles, axions and 
intermediate mass dark matter}, {\it Phys. Lett.} B 
{\bf 489} (2000) 194.

\bibitem{superstrings}
M.B.~Green and J.H.~Schwarz, {\it Anomaly Cancellation in 
Supersymmetric $D=10$ Gauge Theory and Superstring Theory},
{\it Phys. Lett.} B {\bf 149} (1984) 117;
D.J.~Gross, J.A.~Harvey, E.J.~Martinec, and R.~Rohm,
{\it The Heterotic String}, {\it Phys. Rev. Lett.} {\bf 54} 
(1985) 502;
M.~Dine, V.~Kaplunovsky, M.L.~Mangano, C.~Nappi, and 
N.~Seiberg, {\it Superstring Model Building}, {\it Nucl. Phys.} 
B {\bf 259} (1985) 549;
E.~Witten, {\it Symmetry Breaking Patterns in Superstring 
Models}, {\it Nucl. Phys.} B {\bf 258} (1985) 75;
P.~Candelas, G.T.~Horowitz, A.~Strominger, and E.~Witten,
{\it Vacuum Configurations for Superstrings}, {\it Nucl. 
Phys.} B {\bf 258} (1985) 46.

\bibitem{initial}
D.S.~Goldwirth and T.~Piran, {\it Initial conditions for 
inflation}, {\it Phys. Reports} {\bf 214} (1992) 223.

\bibitem{kl} D.A.~Kirzhnits and A.D.~Linde, 	
{\it Macroscopic Consequences of the Weinberg Model},
\plb{42}{471}{1972}.

\bibitem{trinification}
S.L.~Glashow, {\it Trinification of All Elementary Particle 
Forces}, in {\it Fifth Workshop on Grand Unification}, 
edited by K.~Kang, H.M.~Fried, and P.H.~Frampton, World 
Scienific, Singapore 1984; 
K.S.~Babu, X.-G.~He, and S.~Pakvasa, {\it Neutrino Masses 
and Proton Decay Modes in $SU(3)\times SU(3)\times SU(3)$ 
Trinification}, {\it Phys. Rev.} D {\bf 33} (1986) 763;
G.~Lazarides, C.~Panagiotakopoulos, {\it MSSM from SUSY 
trinification}, {\it Phys. Lett.} B {\bf 336} (1994) 190;
S.~Willenbrock, {\it Triplicated trinification}, {\it 
Phys. Lett.} B {\bf 561} (2003) 130;
J.~Sayre, S.~Wiesenfeldt, and S.~Willenbrock,	
{\it Minimal trinification}, {\it Phys. Rev.} D {\bf 73} 
(2006) 035013.

\bibitem{bau} M.~Yoshimura, {\it Unified Gauge Theories 
and the Baryon Number of the Universe}, 
\prl{41}{281}{1978}, (E) {\bf 42} (1979) 746;
A.Yu~Ignatiev, N.V.~Krasnikov, V.A.~Kuzmin, and 
A.N.~Tavkhelidze, {\it Universal CP Noninvariant Superweak 
Interaction and Baryon Asymmetry of the Universe}, 
\plb{76}{436}{1978};
J.R.~Ellis, M.K.~Gaillard, and D.V.~Nanopoulos, {\it 
Baryon Number Generation in Grand Unified Theories},
\plb{80}{360}{1978}, (E) {\bf 82} (1979) 464;
D.~Toussaint, S.B.~Treiman, F.~Wilczek, and A.~Zee, {\it
Matter-Antimatter Accounting, Thermodynamics, and 
Black Hole Radiation},
\prd{19}{1036}{1979};
S.~Weinberg, {\it Cosmological Production of Baryons}, 
\prl{42}{850}{1979}.

\bibitem{kibble} T.W.B.~Kibble, {\it Topology of Cosmic 
Domains and Strings}, 
\jpa{9}{1387}{1976};
T.W.B.~Kibble, {\it Some implications of a cosmological 
phase transition}, \pr{67}{183}{1980}.

\bibitem{string} H.B.~Nielsen and P.~Olesen, {\it Vortex 
Line Models for Dual Strings}, 
\npb{61}{45}{1973}.

\bibitem{wall} Ya.B.~Zeldovich, I.Yu.~Kobzarev, and
L.B.~Okun, {\it Cosmological Consequences of the 
Spontaneous Breakdown of Discrete Symmetry}, 
\jetpsp{40}{1}{1975}.

\bibitem{thermalinf} G.~Lazarides, C.~Panagiotakopoulos, 
and Q.~Shafi, {\it Baryogenesis and the Gravitino Problem 
in Superstring Models}, \prl{56}{557}{1986};
G.~Lazarides and Q.~Shafi, {\it Anomalous discrete 
symmetries and the domain wall problem}, \npb{392}{61}{1993};
D.H.~Lyth and E.D.~Stewart, {\it Cosmology with a TeV mass 
GUT Higgs}, \prl{75}{201}{1995};
D.H.~Lyth and E.D.~Stewart, {\it Thermal inflation and the 
moduli problem}, \prd{53}{1784}{1996}.

\bibitem{cosmicstring} T.W.B.~Kibble, G.~Lazarides, and 
Q.~Shafi, {\it Strings in $SO(10)$},
\plb{113}{237}{1982}.

\bibitem{gws} 
X.~Siemens, V.~Mandic, and J.~Creighton, {\it 
Gravitational wave stochastic background from cosmic  
(super)strings}, {\it Phys. Rev. Lett.} {\bf 98} (2007) 
111101;
M.R.~DePies and C.J.~Hogan, {\it Stochastic Gravitational 
Wave Background from Light Cosmic Strings}, {\it Phys. Rev.} 
D {\bf 75} (2007) 125006;
S.~Kuroyanagi, K.~Miyamoto, T.~Sekiguchi, K.~Takahashi,   
and J.~Silk, {\it Forecast constraints on cosmic string 
parameters from gravitational wave direct detection 
experiments}, {\it Phys. Rev.} D {\bf 86} (2012) 023503;
J.J.~Blanco-Pillado and K.D.~Olum, {\it Stochastic  
gravitational wave background  from smoothed cosmic 
string loops}, {\it Phys. Rev.} D {\bf 96} (2017) 104046;
J.J.~Blanco-Pillado, K.D.~Olum, and X.~Siemens, {\it 
New limits on cosmic strings from gravitational wave 
observation}, {\it Phys. Lett.} B {\bf 778} (2018) 392.

\bibitem{gwsstrings} 
G.~Lazarides and C.~Panagiotakopoulos, {\it Gravitational 
Waves from Double Hybrid Inflation}, {\it Phys. Rev.} D 
{\bf 92} (2015) 123502.

\bibitem{stringsmono}
G.~Lazarides and Q.~Shafi, {\it Monopoles, Strings, and 
Necklaces in $SO(10)$ and $E_6$}, \jhep{10}{193}{2019}. 

\bibitem{axion} G.~Lazarides and Q.~Shafi, {\it Axion 
Models with No Domain Wall Problem},
\plb{115}{21}{1982}.

\bibitem{wallsbounded} T.W.B.~Kibble, G.~Lazarides, and 
Q.~Shafi, {\it Walls Bounded by Strings},
\prd{26}{435}{1982}.

\bibitem{stringsbounded} G.~Lazarides, Q.~Shafi, and 
T.F.~Walsh, {\it Cosmic Strings and Domains in Unified 
Theories}, \npb{195}{157}{1982}.

\bibitem{peebles} P.J.E.~ Peebles, D.N.~Schramm, E.L.~Turner, 
and R.N.~Kron, {\it The Case for the hot big bang cosmology}, 
\n{352}{769}{1991}.

\bibitem{deuterium} S.~Burles, K.M.~Nollett, and M.S.~Turner, 
{\it Big bang nucleosynthesis predictions for precision 
cosmology}, \apj{552}{1}{2001}.

\bibitem{einhorn} M.B.~Einhorn, {\it The Production of 
Magnetic Monopoles in the Very Early Universe}, in 
proceedings of {\it Unification of the Fundamental Particle 
Interactions, Erice 1980}, Editors S.~Ferrara et al, Plenum 
Press, New York 1980.

\bibitem{hz} E.R.~Harrison, {\it Fluctuations at the 
threshold of classical cosmology}, \prd{1}{2726}{1970};
Ya.B.~Zeldovich, {\it A Hypothesis unifying the structure 
and the entropy of the universe}, \mnras{160}{1}{1972}.

\bibitem{extended}
G.~Lazarides and Q.~Shafi, {\it Extended Structures at 
Intermediate Scales in an Inflationary Cosmology}, 
{\it Phys. Lett.} B {\bf 148} (1984) 35;
G.~Lazarides, C.~Panagiotakopoulos and Q.~Shafi, {\it 
Magnetic monopoles from superstring models}, 
\prl{58}{1707}{1987}.

\bibitem{width} A.D.~Dolgov and  A.D.~Linde, {\it 
Baryon Asymmetry in Inflationary Universe},
\plb{116}{329}{1982};
L.F.~Abbott, E.~Farhi, and M.B.~Wise, {\it Particle 
Production in the New Inflationary Cosmology},
\plb{117}{29}{1982}.

\bibitem{dh} A.D.~Dolgov and S.H.~Hansen, {\it Equation 
of motion of a classical scalar field with back 
reaction of produced particles},
\npb{548}{408}{1999}.

\bibitem{slowroll} P.J.~Steinhardt and M.S.~Turner, {\it 
A Prescription for Successful New Inflation},
\prd{29}{2162}{1984}.

\bibitem{oscillation} M.S.~Turner, {\it Coherent Scalar 
Field Oscillations in an Expanding Universe}, 
\prd{28}{1243}{1983}.

\bibitem{hawking} S.W.~Hawking, {\it Black hole 
explosions?}, \n{248}{30}{1974};
S.W.~Hawking, {\it Particle Creation by Black Holes}, 
\cmp{43}{199}{1975}, (E) {\bf 46} (1976) 206;
G.W.~Gibbons and S.W.~Hawking, {\it 
Action Integrals and Partition Functions in Quantum 
Gravity}, \prd{15}{2752}{1977}.

\bibitem{bunch} T.~Bunch and P.C.W.~Davies, {\it Quantum 
field theory in de Sitter space: Renormalization by 
point-splitting}, \prsla{360}{117}{1978};
A.~Vilenkin and L.H.~Ford,
{\it Gravitational Effects upon Cosmological Phase 
Transitions}, \prd{26}{1231}{1982};
A.D.~Linde, {\it Scalar Field Fluctuations in 
Expanding Universe and the New Inflationary Universe 
Scenario}, \plb{116}{335}{1982};
A.A.~Starobinsky, {\it Dynamics of Phase Transition 
in the New Inflationary Universe Scenario and Generation 
of Perturbations}, \plb{117}{175}{1982}.

\bibitem{fischler} W.~Fischler, B.~Ratra, and
L.~Susskind, {\it Quantum Mechanics of Inflation}, 
\npb{259}{730}{1985}, (E) {\bf 268} (1986) 747.

\bibitem{zeta} J.M.~Bardeen, P.J.~Steinhardt, and
M.S.~Turner, {\it Spontaneous Creation of Almost 
Scale-Free Density Perturbations in an Inflationary 
Universe}, \prd{28}{679}{1983}.

\bibitem{zetarev} V.F.~Mukhanov, H.A.~Feldman, and
R.H.~Brandenberger, {\it Theory of cosmological 
perturbations. Part 1. Classical perturbations. Part 2. 
Quantum theory of perturbations. Part 3. Extensions}, 
\pr{215}{203}{1992}.

\bibitem{sachswolfe} R.K.~Sachs and A.M.~ Wolfe,
{\it Perturbations of a cosmological model and angular 
variations of the microwave background},
\apj{147}{73}{1967}.

\bibitem{tensor} V.A.~Rubakov, M.V.~Sazhin, and
A.V.~Veryaskin, {\it Graviton Creation in the 
Inflationary Universe and the Grand Unification 
Scale}, \plb{115}{189}{1982};
R.~Fabbri and M.D.~Pollock, {\it The Effect of 
Primordially Produced Gravitons upon the Anisotropy 
of the Cosmological Microwave Background Radiation}, 
\plb{125}{445}{1983};
L.F.~Abbott and M.B.~Wise, {\it Constraints on 
Generalized Inflationary Cosmologies}, 
\npb{244}{541}{1984};
B.~Allen, {\it The Stochastic Gravity Wave Background 
in Inflationary Universe Models}, \prd{37}{2078}{1988};
M.J.~White, {\it Contribution of long wavelength 
gravitational waves to the cosmic microwave background 
anisotropy}, \prd{46}{4198}{1992}.

\bibitem{vlachos}
G.~Lazarides, Q.~Shafi, and N.D.~Vlachos, {\it 
Supersymmetric inflation, baryogenesis and 
$\nu_\mu-\nu_\tau$ oscillations}, {\it Phys. Lett.} 
B {\bf 427} (1998) 53;
G.~Lazarides and N.D.~Vlachos, {\it Hierarchical neutrinos 
and supersymmetric inflation}, {\it Phys. Lett.} B {\bf 459} 
(1999) 482;
G.~Lazarides, {\it Supersymmetric hybrid inflation}, {\it 
NATO Sci. Ser. II} {\bf 34} (2001) 399 [hep-ph/0011130].

\bibitem{pedestrians} 	
W.~Buchm\"{u}ller, P.~Di Bari, and M.~Pl\"{u}macher, {\it 
Leptogenesis for pedestrians}, {\it Annals Phys.} {\bf 315} 
(2005) 305.

\bibitem{vacuum} C.~Callan, R.~Dashen, and D.~Gross, {\it 
The Structure of the Gauge Theory Vacuum},
\plb{63}{334}{1976};
R.~Jackiw and C.~Rebbi, {\it Vacuum Periodicity in a 
Yang-Mills Quantum Theory}, \prl{37}{172}{1976}.

\bibitem{sphaleronsol} N.S.~Manton, {\it Topology in the 
Weinberg-Salam Theory}, \prd{28}{2019}{1983};
F.R.~Klinkhamer and N.S.~Manton, {\it A Saddle Point 
Solution in the Weinberg-Salam Theory}, 
\prd{30}{2212}{1984}.

\bibitem{thooft} G.'t Hooft, {\it Symmetry Breaking Through 
Bell-Jackiw Anomalies}, \prl{37}{8}{1976};
G.'t Hooft, {\it Computation of the Quantum Effects Due to 
a Four-Dimensional Pseudoparticle}, \prd{14}{3432}{1976}.

\bibitem{turneribanez} J.A.~Harvey and M.S.~Turner, {\it 
Cosmological baryon and lepton number in the presence of 
electroweak fermion number violation}, \prd{42}{3344}{1990};
L.E.~Ib\'a\~nez and F.~Quevedo, {\it 
Supersymmetry protects the primordial baryon asymmetry}, 
\plb{283}{261}{1992}.

\bibitem{supernovae}
P.M. Garnavich et al., {\it Supernova limits on the cosmic 
equation of state}, {\it Astrophys. J.} {\bf 509} (1998) 74;
A.G.~Riess et al., {\it Observational Evidence from 
Supernovae for an Accelerating Universe and a Cosmological 
Constant}, {\it Astron. J.} {\bf 116} (1998) 1009; 
S.~Perlmutter et al., {\it Measurements of $\Omega$ and 
$\Lambda$ from 42 high redshift supernovae}, {\it Astrophys. 
J.} {\bf 517} (1999) 565.

\bibitem{quintessence} 
R.R.~Caldwellet,  R.~Dave, and P.J.~Steinhardt, {\it 
Cosmological imprint of an energy component with general 
equation of state}, {\it Phys. Rev. Lett.} {\bf 80} (1998) 
1582.

\bibitem{quintreview} 
P.~Binetrui, {\it Cosmological constant versus 
quintessence}, {\it Int. J. Theor. Phys.} {\bf 39} (2000) 1859;
E.J.~Copeland, M.~Sami, and S.~Tsujikawa, {\it Dynamics 
of dark energy}, {\it Int. J. Mod. Phys.} D {\bf 15} (2006) 
1936.

\bibitem{LSPrelic}
M.E.~Gomez, G.~Lazarides, and C.~Pallis, {\it Supersymmetric 
cold dark matter with Yukawa unification}, {\it Phys. Rev.} D 
{\bf 61} (2000) 123512;
J.R.~Ellis, T.~Falk, K.A.~Olive, and M.~Srednicki, {\it
Calculations of neutralino-stau coannihilation channels and 
the cosmologically relevant region of MSSM parameter space},
{\it Astropart. Phys.} {\bf 13} (2000) 181, (E) {\bf 15} (2001) 
413.

\bibitem{micromegas}
G.~Belanger, F.~Boudjema, A.~Pukhov, and A.~Semenov,
{\it $micrOMEGAs_3$: A program for calculating dark matter 
observables}, {\it Comput. Phys. Commun.} {\bf 185} (2014) 960.

\bibitem{axionDM}
J.~Preskill, M.B.~Wise, and F.~Wilczek, {\it Cosmology of 
the Invisible Axion}, {\it Phys. Lett.} B {\bf 120} (1983) 127;
L.F.~Abbott and P.~Sikivie, {\it A Cosmological Bound on 
the Invisible Axion}, {\it Phys. Lett.} B {\bf 120} (1983) 133;
M.~Dine and W.~Fischler, {\it The Not So Harmless Axion}, 
{\it Phys. Lett.} B {\bf 120} (1983) 137.

\bibitem{iso}
M.~Kawasaki, E.~Sonomoto, and T.T.~Yanagida, {\it 
Cosmologically allowed regions for the axion decay 
constant $F_a$}, {\it Phys. Lett.} B {\bf 782} (2018) 181 
and references therein.

\bibitem{axionmuterm}
G.~Lazarides, C.~Panagiotakopoulos, and Q.~Shafi,
{\it Axion, $\mu$ term, and supersymmetric hybrid 
inflation}, {\it Phys. Rev.} D {\bf 95} (2017) 055017.

\bibitem{XENON}
E. Aprile et al., {\it Dark Matter Search Results from a 
One Ton-Year Exposure of XENON1T}, {\it Phys. Rev. Lett.} 
{\bf 121} (2018) 111302.

\bibitem{gravitinoDM}
J.R.~Ellis, K.A.~Olive, Y.~Santoso, and V.C.~Spanos, {\it 
Gravitino dark matter in the CMSSM}, {\it Phys. Lett.} B 
{\bf 588} (2004) 7;
K.~Jedamzik, K.-Y.~Choi, L.~Roszkowski, and R.~Ruiz de Austri,
{\it Solving the cosmic lithium problems with gravitino dark 
matter in the CMSSM}, {\it J. Cosmol. Astropart. Phys.} {\bf 07} 
(2006) 007;
J.~Pradler and F.D.~Steffen, {\it Implications of Catalyzed 
BBN in the CMSSM with Gravitino Dark Matter}, {\it Phys. Lett.} B 
{\bf 666} (2008) 181.

\bibitem{axino}
L.~Covi, L.~Roszkowski, R.~Ruiz de Austri, and M.~Small,
{\it Axino dark matter and the CMSSM},
{\it J. High Energy Phys.} {\bf 06} (2004) 003;
H.~Baer, A.D.~Box, and H.~Summy, {\it Neutralino versus 
axion/axino cold dark matter in the 19 parameter SUGRA 
model}, {\it J. High Energy Phys.} {\bf 10} (2010) 023;
K.-Y.~Choi, L.~Covi, J.E.~Kim, and L.~Roszkowski, {\it
Axino Cold Dark Matter Revisited}, {\it J. High Energy Phys.} 
{\bf 04} (2012) 106.

\bibitem{renorm}
E.S.~Abers and B.W.~Lee, {\it Gauge Theories}, 
{\it Phys. Reports} {\bf 9} (1973) 1 and references therein.

\bibitem{finite}
J.H.~Schwarz, {\it Superstring theory}, {\it Phys. Reports}
{\bf 89} (1982) 223.

\bibitem{landscape}
L.~Susskind, {\it The Anthropic Landscape of String Theory},
arXiv:hep-th/0302219.

\bibitem{classification}
N.~Ganoulis, G.~Lazarides, and Q. Shafi, {\it 
Classification of Three Generation Superstring Models 
According to Their Discrete Symmetries},
{\it Nucl. Phys.} B {\bf 323} (1989) 374.

\bibitem{mtheory}
E.~Witten, {\it Five-branes and M theory on an orbifold},
{\it Nucl. Phys.} B {\bf 463} (1996) 383.

\bibitem{initsuper}    
G.~Lazarides and N.D.~Vlachos, {\it Initial conditions for 
supersymmetric inflation}, {\it Phys. Rev.} D {\bf 56} (1997) 
4562.

\bibitem{preBB}
M.~Gasperini and G.~Veneziano, {\it Pre-big bang in string 
cosmology}, {\it Astropart. Phys.} {\bf 1} (1993) 317.

\bibitem{2stage}
G.~Lazarides and N.~Tetradis, {\it Two stage inflation in 
supergravity}, {\it Phys. Rev.} D {\bf 58} (1998) 123502;
C.~Panagiotakopoulos and N.~Tetradis, {\it Two stage 
inflation as a solution to the initial condition problem 
of hybrid inflation}, {\it Phys. Rev.} D {\bf 59} (1999) 
083502.

\bibitem{bicep2}
P.A.R.~Ade et al., {\it Detection of $\bf{\bar{B}}$-Mode 
Polarization at Degree Angular Scales by BICEP2},
{\it Phys. Rev. Lett.} {\bf 112} (2014) 241101.

\end{thebibliography}
\end{document}